# Magneto-photoluminescence measurements of symmetric and asymmetric CdSe/ZnSe self-assembled quantum dots

**K. P. Hewaparakrama, N. Mukolobwiez , L. M. Smith, H. E. Jackson**
Dept. of Physics, Univ. of Cincinnati, Cincinnati
**S. Lee, M. Dobrowolska, J. Furdyna**
Dept. of Physics, Univ. of Notre Dame

**Abstract.** We study the excitonic structure of CdSe/ZnSe self assembled quantum dots (SAQD) by magneto-photoluminescence (PL) spectroscopy. Using fixed 200 nm apertures through a metal film, we are able to probe *single* narrow (200 μeV) spectroscopic lines emitted from CdSe quantum dots at 2K. Using linear polarization analysis of these lines, we find that approximately half of the quantum dots are elliptically elongated along the [110] direction. The other half of the QDs are symmetric, with emission lines which are completely unpolarized at zero field and exhibit no doublet structure. Using an applied magnetic field, we obtain the diamagnetic shift and the g-factor for several symmetric dots and find that the variation from dot to dot is random, with no systematic dependence with emission energy. In addition, we find no correlation between the diamagnetic shift and the g-factor.

## Introduction

In the last two decades, significant effort to study lower dimensional semiconductor nano-structures has been carried out. Self Assembled Quantum Dots (SAQDs) are formed naturally through a three dimensional growth mode driven by the larg (typically 7%) lattice mismatch between two different semiconductors: InAs on GaAs, CdSe on ZnSe, or SiGe on Si. SAQD structures strongly confine electrons and holes as demonstrated by the very narrow PL emission lines seen in these structures. However, understanding the internal energy structure of these naturally formed quantum dots is often quite difficult because of the random variation of the ensemble of dots.

In this paper, we study the magneto-PL of CdSe SAQDs to probe their internal structure. Several theoretical papers have suggested that the diamagnetic shift and the spin splitting are quite sensitive to the size and geometry of the quantum dots [1, 2]. Using an aperture defined through a metal mask, we study a distribution of nearly 100 dots. Using linearly polarized PL, we identified elliptical and cylindrically symmetric dots. Then we obtained the diamagnetic shift and g-factor for the nearly 50 symmetric dots which exhibited no noticeable polarization or splitting.

**Experimental Details**

The sample was grown by MBE on a GaAs substrate. After a 1 μm ZnSe buffer, about 3 monolayers of CdSe were deposited followed by a thin ZnSe capping layer. In order to select the small number of dots, an aluminium pad with 16 apertures from 6 μm down to 100 nm in diameter were prepared by electron beam lithography. The sample was maintained at 2K in an optical cryostat, excited by the 514 nm line of cw Ar ion laser with power of ~100 μW. The emitted PL was dispersed by a 1 meter focal length spectrometer and detected by a cooled CCD camera. Magnetic fields up to 3 T were applied to the sample by using a small superconducting solenoid coil, and polarization of the emitted PL was determined using a Glan-Thomson linear polarizer and Babinet-Soleil compensator. Figure 1 shows the photoluminescence from CdSe SAQD collected through the 100 nm diameter aperture. For this aperture one can clearly see individual sharp spectral peaks.

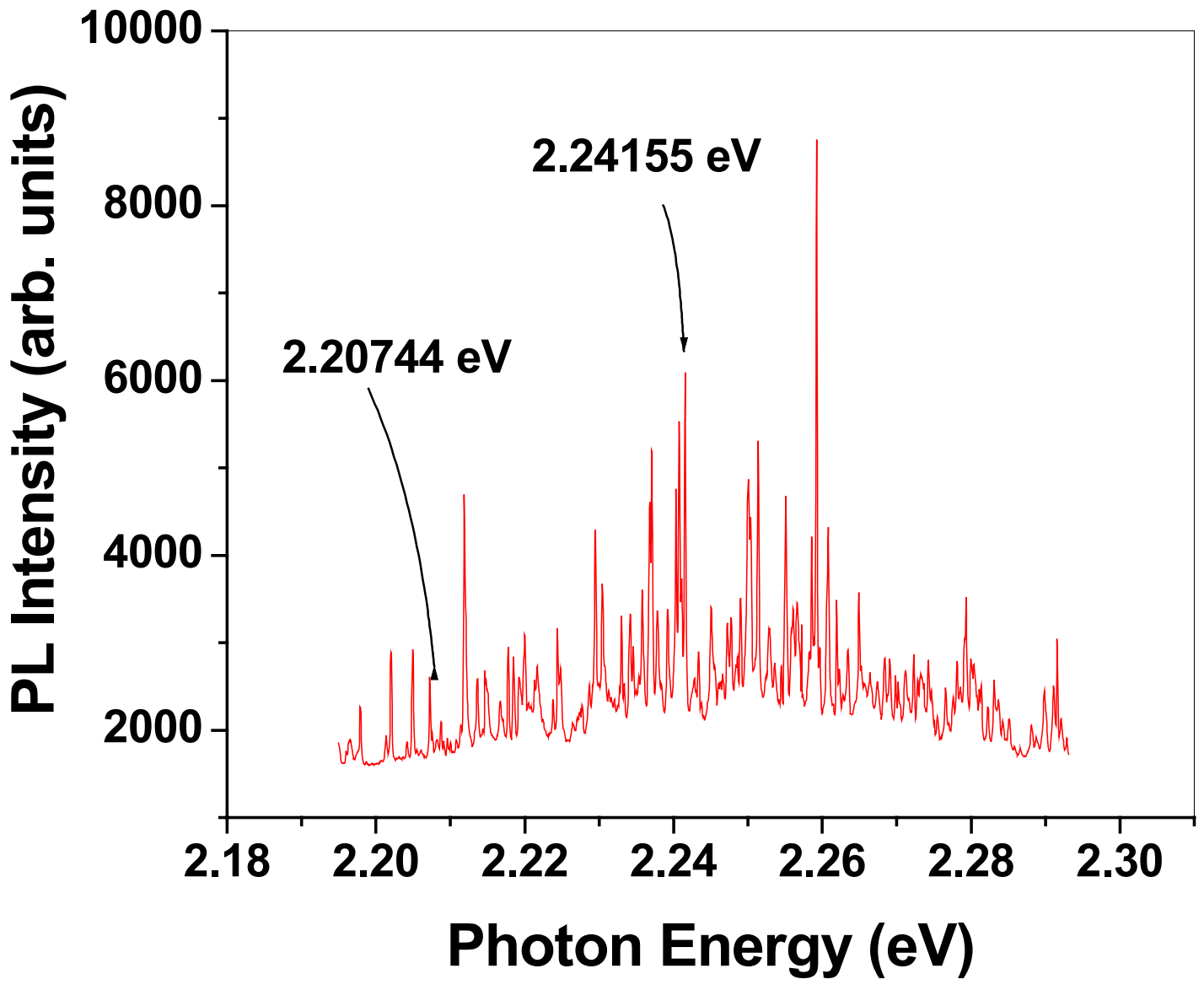

**Figure 1**: PL spectrum from CdSe SAQDs through a 100 nm diameter aperture.

**Results and discussion**.

Figures 2 and 3 show linearly-polarized PL spectra for two different quantum dots (noted by arrows in Fig. 1) in zero magnetic field. As seen in Fig. 2, the exciton emission from one QD consists of two components with orthogonal linear polarizations along the [110] and [1,-1,0] crystal orientations. The energy splitting between the two peaks is about 340 μeV. In contrast, the other QD shown in Fig. 3 exhibits no linear polarization of the emission. Such behavior has been seen before in CdSe SAQDs [3, 4] and reflects the internal symmetry of the quantum dots. The linearly-polarized doublets occur because the QD is elliptically deformed along the [110] crystallographic directions. On the other hand, for dots which are in-plane symmetric, there is no preferred orientation and so one observes an unpolarized emission line. Figure 4 shows a statistical histogram of the zero-field splitting for 50 of the SAQD emission lines. We find that half of the SAQDs sampled are symmetric, while the remainder are elongated along the [110] crystallographic direction. A small number appear to be aligned along the [1-10] crystallographic direction.

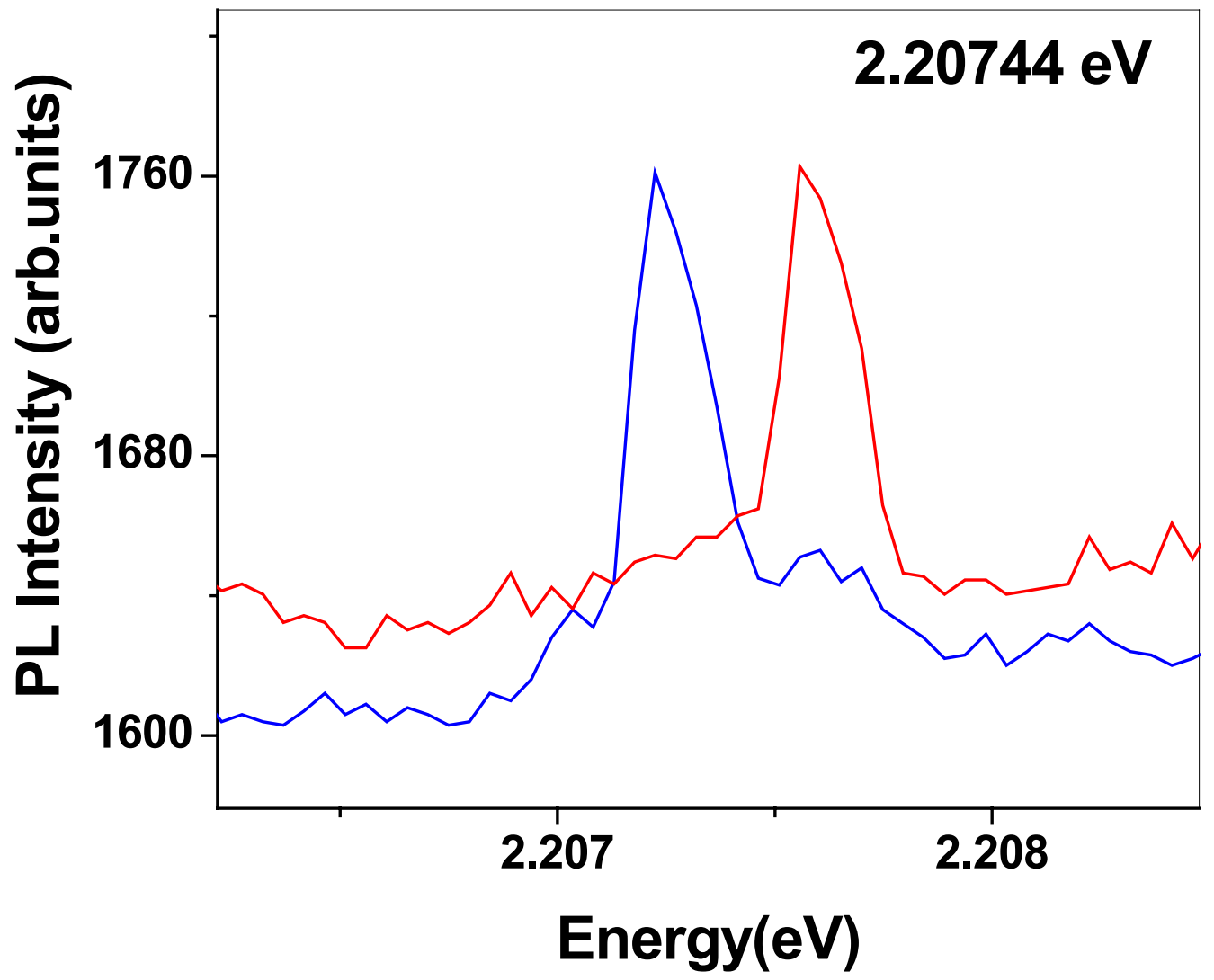

**Figure 2**: Linearly-polarized photoluminescence spectra from an asymmetric single QD at zero magnetic field at 2.20744 eV.

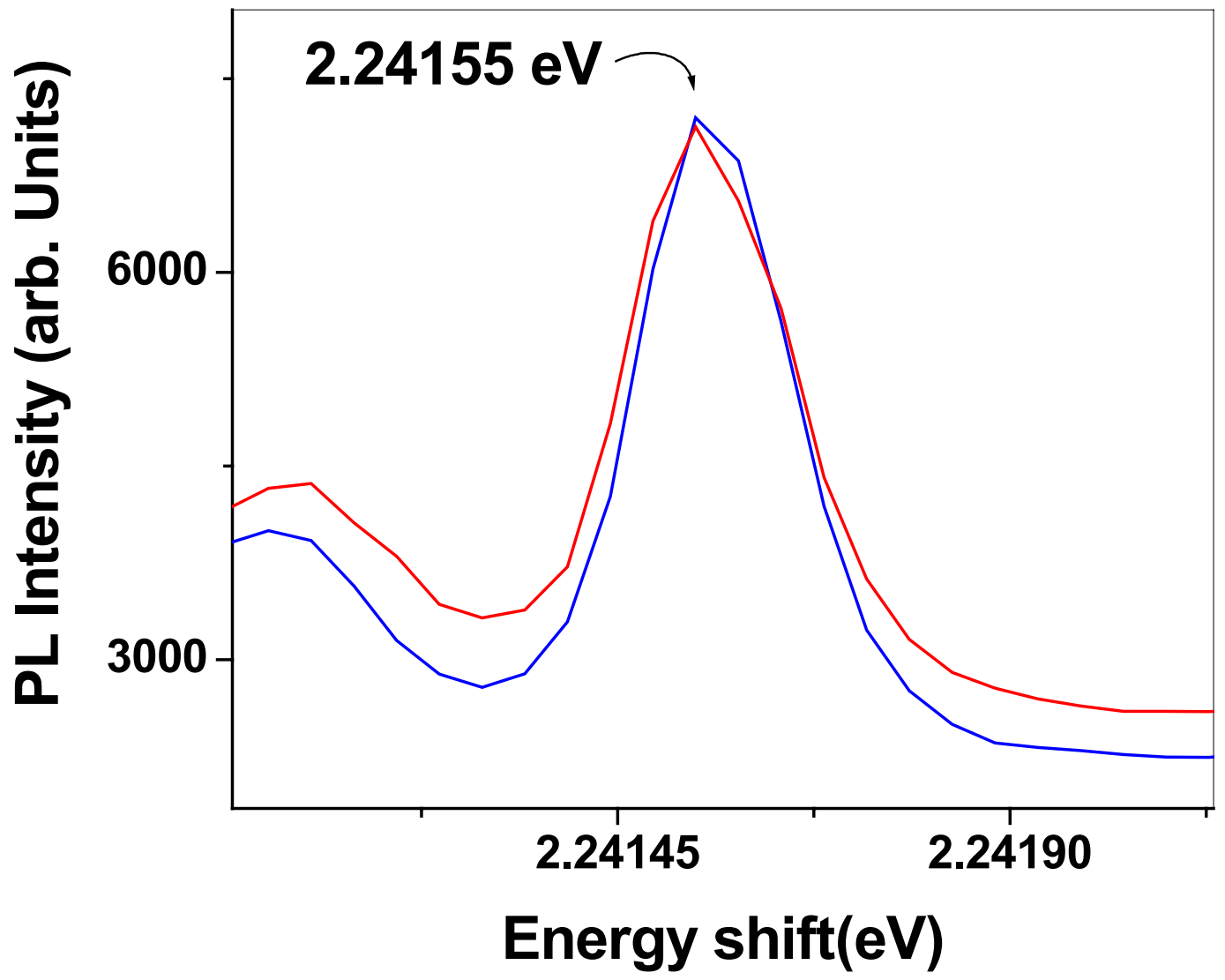

**Figure 3:** Linearly-polarized nano-PL spectra from a symmetric single QD in zero magnetic field at 2.24155 eV.

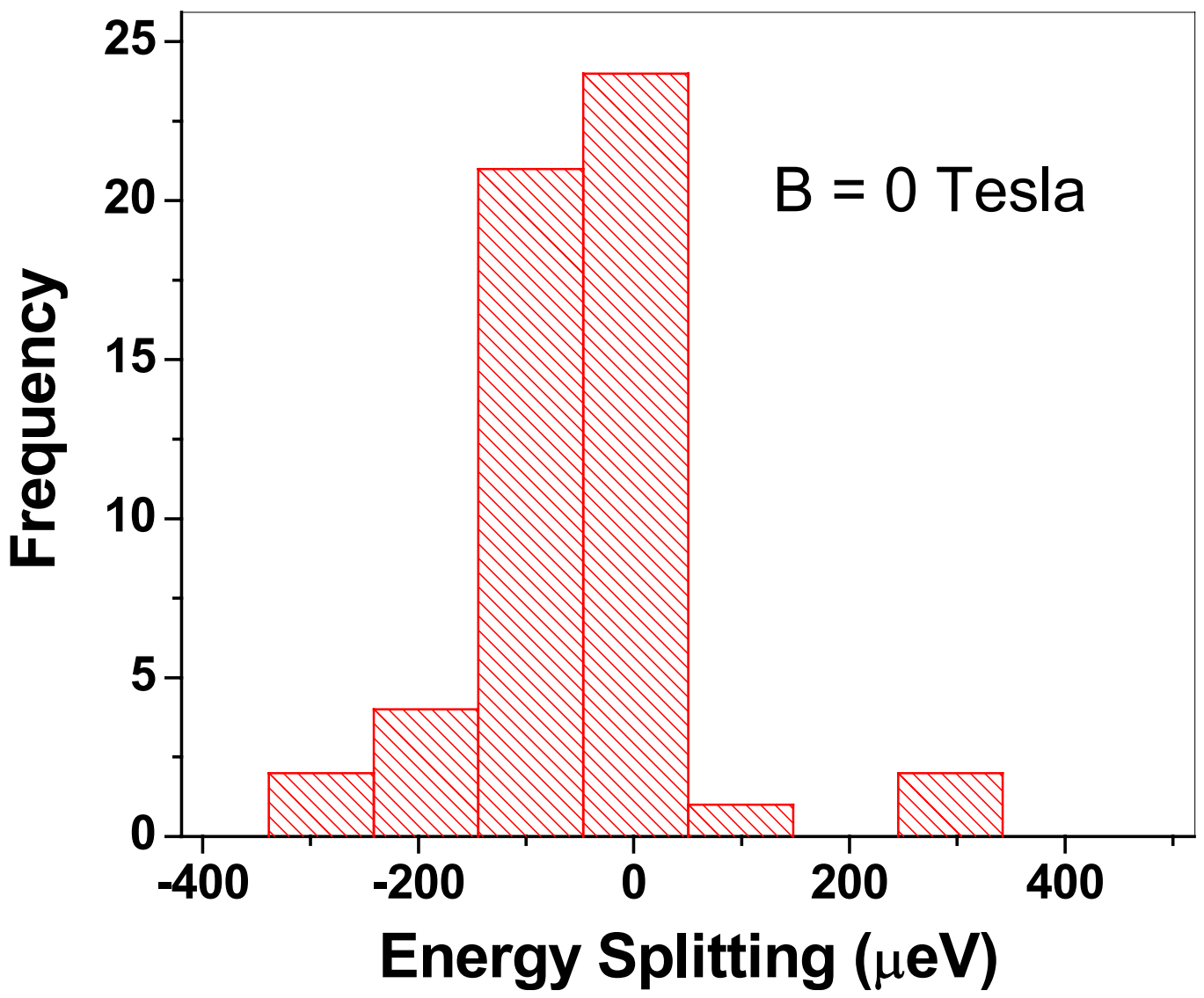

**Figure 4**: Statistical histogram of the zero-field splitting for 50 of the SAQD emission lines.

For the symmetric dots, we have made nano-magneto-PL measurements using a small superconducting solenoid with an integrated microscope objective. These measurements show that the single lines split (see Fig. 5) into two circularly polarized lines (σ+ and σ-) which are emitted from decay of the spin-polarized confined excitons (note that only the excitons |h,e> = |+3/2,+1/2> and |-3/2,-1/2>, ΔJ = +1 or –1 are optically active). The fan-diagram which displays the energy position of each line as a function of magnetic field is shown in Fig. 6. The energy position of each state is modelled using $\Delta E = \gamma B^2 \pm \mu_B g^* B$. By fitting the magneto-PL data, we can extract the diamagnetic shift parameter, γ, and the effective g-factor, g* for each line.

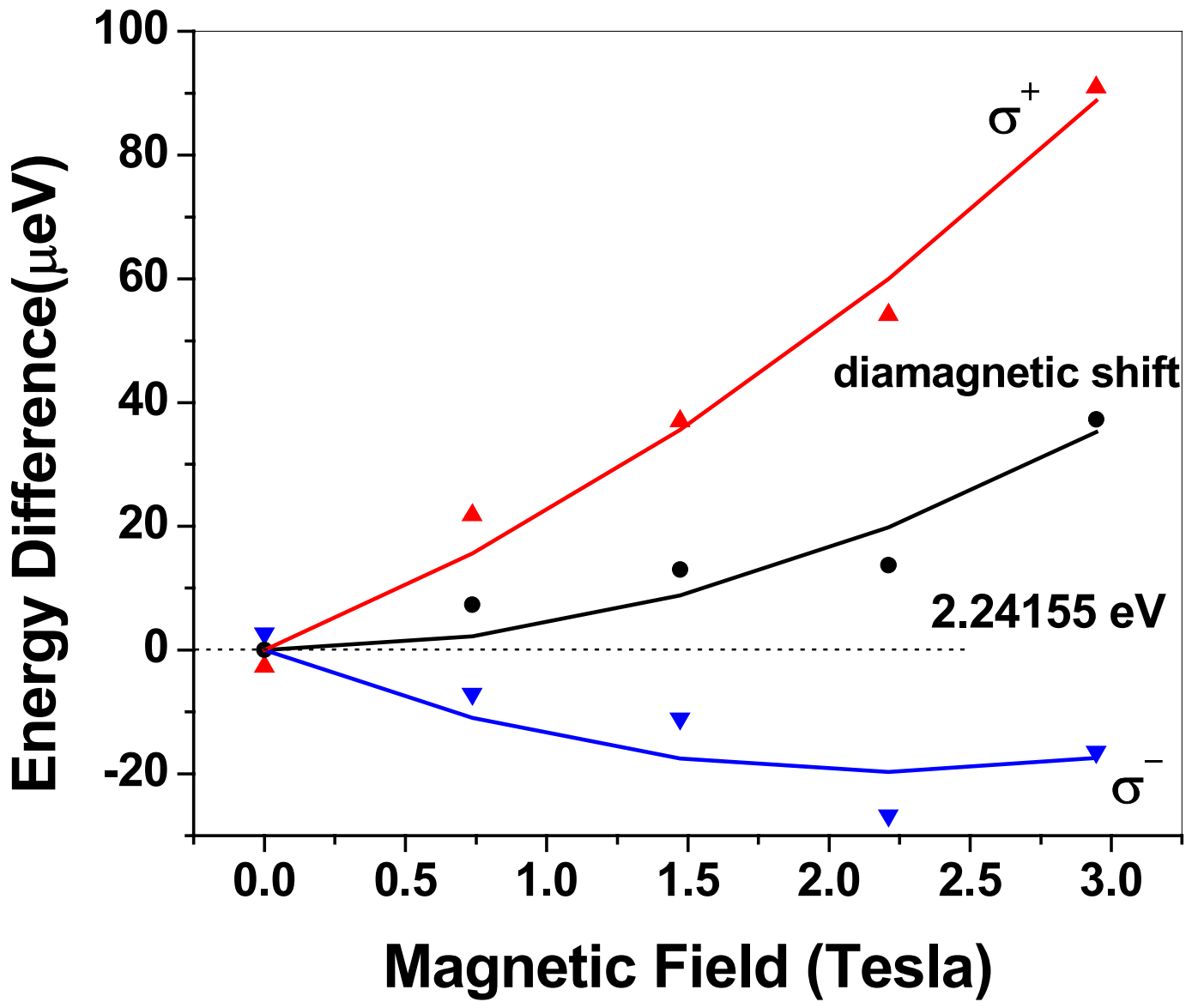

**Figure 5**: Magnetic field dependence of exciton energy for the symmetric QD at 2.24156 eV.

We find that the $\gamma$ values range from 2 to 10 $\mu eV/T^2$ with an average value of 6 $\mu eV/T^2$, and the g* values range from 0.2 to 1.2 with an average value of 0.7. The values of these parameters show no dependence on the emission energy. Moreover, we find no correlation between the values of $\gamma$ or g*. Because one would expect that $\gamma$ and g* should vary with size, this result suggests that the dominant variation in emission energy is not in size fluctuations, but rather alloy disorder. On the other hand, the value of these parameters is significantly different than has been observed previously on CdSe dots grown by two other groups [3, 4]. Both of these groups measured $\gamma$ to be 0.7 $meV/T^2$ and g* to be 1.6. These values were obtained from single dots on small-etched mesas. This variation may also result from size and alloy differences.

**Conclusion**

We have used polarized magneto-photoluminescence to study the symmetry and exciton structure of *single* self-assembled CdSe quantum dots. In zero-field we show that the ensemble of quantum dots can be separated into two distinct types: (1) those dots which are exhibit closely spaced doublets which are orthogonally polarized and (2) those dots which exhibit unpolarized emission lines. We conclude that those dots which exhibit polarized doublets are elliptically elongated along the [110] crystal axis, while those dots which show unpolarized emission lines are symmetric. Finally, we have measured the behavior of excitons confined to twelve symmetric dots in a magnetic field and have extracted their diamagnetic shift and spin splitting. These results show a statistical variation around effective g-factors of 0.7 and $\gamma$ of 6 $\mu eV/T^2$. This variation is not correlated with the emission energy. Moreover, no correlation between g* and $\gamma$, is observed.

We gratefully acknowledge the financial support of the National Science Foundation through grants NSF DMR 0071797, 9975655 and 0072897 and the DARPA-SPINS Program.